\documentclass[journal]{IEEEtran}
\pdfoutput=1

\usepackage{subcaption}
\usepackage{multirow}
\usepackage{booktabs}
\usepackage{stfloats}

\usepackage{cite}

\ifCLASSINFOpdf
   \usepackage[pdftex]{graphicx}
\else
   \usepackage[dvips]{graphicx}
\fi
\usepackage{amsmath}
\usepackage{amssymb}

\begin{document}
%
\title{Learning for Video Compression}
%
%
%

\author{
	Zhibo~Chen,~\IEEEmembership{Senior~Member,~IEEE,}
	~Tianyu He,~Xin Jin,~Feng Wu,~\IEEEmembership{Fellow,~IEEE}
	\thanks{This work was supported in part by the National Key Research and Development Program of China under Grant No. 2016YFC0801001, the National Program on Key Basic Research Projects (973 Program) under Grant 2015CB351803, NSFC under Grant 61571413, 61632001, 61390514. (\textit{Corresponding author: Zhibo Chen})}
	\thanks{Zhibo Chen, Tianyu He, Xin Jin and Feng Wu are with the CAS Key Laboratory of Technology in Geo-spatial Information Processing and Application System, University of Science and Technology of China, Hefei 230027, China (e-mail: chenzhibo@ustc.edu.cn; hetianyu@mail.ustc.edu.cn; jinxustc@mail.ustc.edu.cn; fengwu@ustc.edu.cn)}
	\thanks{Copyright \copyright 2018 IEEE. Personal use of this material is permitted. However, permission to use this material for any other purposes must be obtained from the IEEE by sending an email to pubs-permissions@ieee.org.}
}

%
%

\markboth{IEEE Transactions on Circuits and Systems for Video Technology}%
{Shell \MakeLowercase{\textit{et al.}}: Bare Demo of IEEEtran.cls for IEEE Journals}
%



\maketitle

\begin{abstract}
One key challenge to learning-based video compression is that motion predictive coding, a very effective tool for video compression, can hardly be trained into a neural network. In this paper we propose the concept of PixelMotionCNN (PMCNN) which includes motion extension and hybrid prediction networks. PMCNN can model spatiotemporal coherence to effectively perform predictive coding inside the learning network. On the basis of PMCNN, we further explore a learning-based framework for video compression with additional components of iterative analysis/synthesis, binarization, etc.  Experimental results demonstrate the effectiveness of the proposed scheme. Although entropy coding and complex configurations are not employed in this paper, we still demonstrate superior performance compared with MPEG-2 and achieve comparable results with H.264 codec. The proposed learning-based scheme provides a possible new direction to further improve compression efficiency and functionalities of future video coding. 
\end{abstract}


%

\section{Introduction}
\label{introduction}

\IEEEPARstart{V}{ideo} occupies about $75\%$ of the data transmitted on world-wide networks and that percentage has been steadily growing and is projected to continue to grow further \cite{cisco2017cisco}. Meanwhile the introduction of ultra-high definition (UHD), high dynamic range (HDR), wide color gamut (WCG), high frame rate (HFR) and future immersive video services have dramatically increased the challenge. Therefore, the need for highly efficient video compression technologies are always pressing and urgent.

Since the proposal of concept of hybrid coding by Habibi in 1974 \cite{habibi1974hybrid} and hybrid spatial-temporal coding framework by Forchheimer in 1981 \cite{forchheimer1981differential}, this Hybrid Video Coding (HVC) framework has been widely adopted into most popular existing image/video coding standards like JPEG, H.261, MPEG-2, H.264, and H.265, etc. The video coding performance improves around $50\%$ every 10 years under the cost of increased computational complexity and memory. And now it encountered great challenges to further significantly improve the coding efficiency and to deal efficiently with novel sophisticated and intelligent media applications such as face/body recognition, object tracking, image retrieval, etc. So there exist strong requirements to explore new video coding directions and frameworks as potential candidates for future video coding schemes, especially considering the outstanding development of machine learning technologies.

\begin{figure}[!t]
	\begin{subfigure}{0.22\textwidth}
		\includegraphics[width=\linewidth]{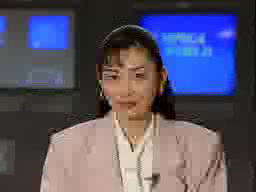}
		\captionsetup{font={footnotesize}}
		\caption{MPEG-2} \label{fig:fig1a}
	\end{subfigure}
	\hspace{0.1cm}
	\begin{subfigure}{0.22\textwidth}
		\includegraphics[width=\linewidth]{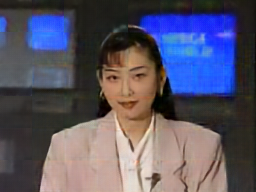}
		\captionsetup{font={footnotesize}}
		\caption{Ours} \label{fig:fig1b}
	\end{subfigure}
	\centering
	\caption[font=footnotesize]{Visualization of reconstructed video compared with MPEG-2 under the compression ratio of about $575$. It is worth noting that entropy coding is not employed for our results, even though it is commonly done in standard video compression codecs.} \label{fig:1}
\end{figure}

Recently, some learning-based image compression schemes have been proposed \cite{toderici2016full,balle2016end,theis2017lossy} with types of neural networks such as auto-encoder, recurrent network and adversarial network, demonstrating a new direction of image/video compression. For example, the first work of learning-based image compression \cite{toderici2015variable,toderici2016full} was introduced in 2016 and demonstrates their better performance compared with the first image coding standard JPEG.

\begin{figure*}[t]
	\centerline{\includegraphics[width=14cm]{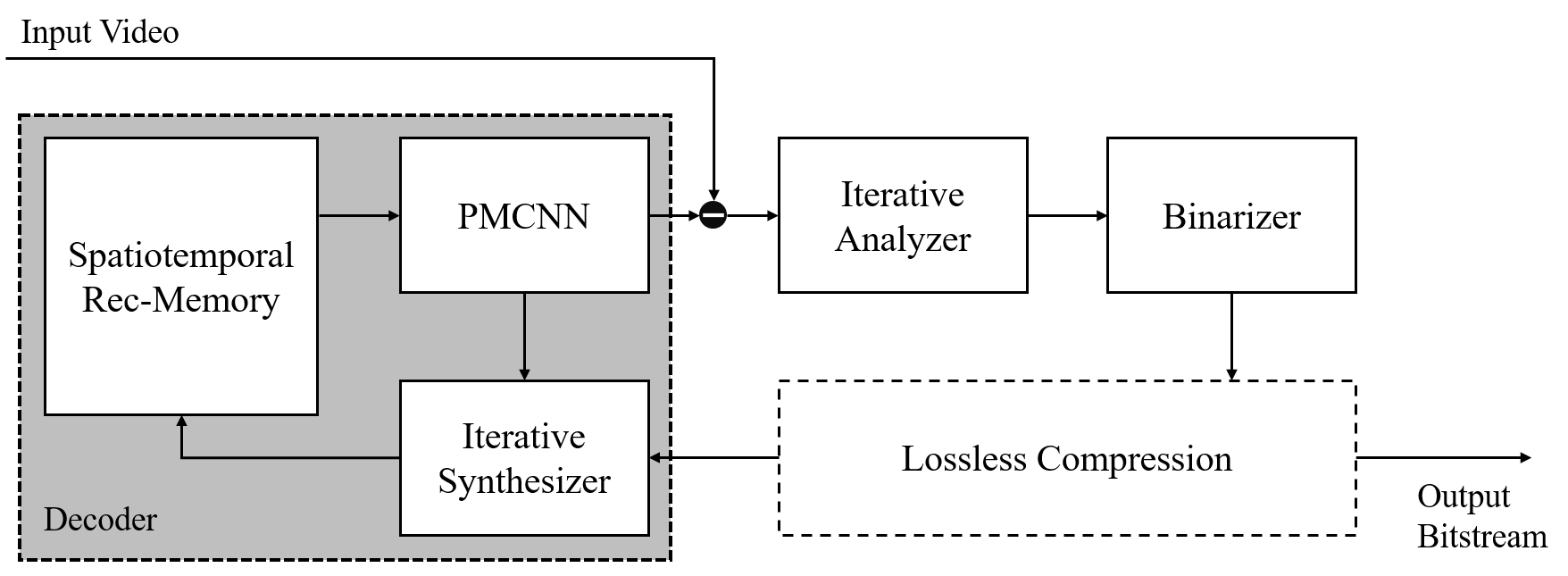}}
	\caption{The pipeline of the proposed learning-based encoder. The encoder converts video into a compressed format (bitstream), then the decoder converts bitstream back into a reconstructed video. Note that, at decoding time, we only have access to the reconstructed data (Spatiotemporal Rec-Memory) instead of original data, therefore, the decoder is included in the encoder to produce reconstructed data for sequentially encoding. We do not employ lossless compression (entropy coding) in this paper, it can be complemented as dashed line in the figure. The details of each components are described in Section \ref{learning}.}
	\centering
	\label{fig:fig2}
\end{figure*}

However, all learning-based methods proposed so far were developed for still image compression and there is still no published work for video compression. One key bottleneck is that motion compensation, as a very effective tool for video coding, can hardly be trained into a neural network (or would be tremendously more complex than conventional motion estimation) \cite{Ohm2017future}. Therefore, there exist some research work on replacing some modules (e.g., sub-pel interpolation, up-sampling filtering, post-processing, etc.) in HVC framework by learning-based modules \cite{prakash2017semantic,yan2017convolutional,jiang2017end}. However, such partial replacements are still under the heuristically optimized HVC framework without capability to successfully deal with aforementioned challenges.

In this paper, we first propose the concept of PixelMotionCNN (PMCNN) by modeling spatiotemporal coherence to effectively deal with the aforementioned bottleneck of motion compensation trained into a neural network, and explore a learning-based framework for video compression. Specifically, we construct a neural network to predict each block of video sequence conditioned on previously reconstructed frame as well as the reconstructed blocks above and to the left of current block. The difference between predicted and original pixels is then analyzed and synthesized iteratively to produce a compact discrete representation. Consequently, the bitstream is obtained that can be used for storage or transmission. To the best of our knowledge, this is the first fully learning-based video compression framework.

The remainder of this paper is organized as follows. Section \ref{relatedwork} introduces the related work. In Section \ref{spatiotemporal}, we explained spatiotemporal modeling with PixelMotionCNN, and the learning-based video compression framework is illustrated in Section \ref{learning}. We will present the experiment results and analysis in Section \ref{experimental}, and then conclude in Section \ref{conclusion}.

\section{Related Work}
\label{relatedwork}
Recently there are two kinds of research work trying to apply machine learning techniques into image/video compression problem, one is Codec-based improvements which introduces learning-based optimization modules combined with traditional image/video codecs, another is pure Learning-based compression framework which are mainly focused on learning-based image compression schemes in current stage. 

\subsection{Codec-based Improvements}

Lossy image/video codecs, such as JPEG and High Efficiency Video Coding (HEVC) \cite{sullivan2012overview}, give a profound impact on image/video compression. Considering the recent success of neural network-based methods in various domains, a stream of CNN-based improvements for these codecs have been proposed to further increase coding efficiency. Most of these works focus on enhancing the performance \cite{yan2017convolutional,song2017neural} or reducing the complexity \cite{li2017deep,yu2015vlsi} of codec by replacing manually designed function with learning-based approach. Similarly, a series of works adopt CNN in post-processing to reduce the artifacts of compression \cite{dong2015compression,wang2017novel,dai2017convolutional}. Encouraged by positive results in domain of super-resolution, another line of work encodes the down-sampled content with codec and then up-samples the decoded one by CNN for reconstruction \cite{jiang2017end,li2017convolutional}.

\subsection{Learning-based Image Compression}

End-to-end image compression has surged for almost two years, opening up a new avenue for lossy compression. The majority of them adopt an autoencoder-like scheme.

In the works of \cite{gregor2016towards,toderici2015variable,toderici2016full, johnston2017improved}, a discrete and compact representation is obtained by applying a quantization to the bottleneck of auto-encoder. To achieve variable bit rates, the model progressively analyzes and synthesizes residual errors with several auto-encoders. Progressive codes are essential to Rate-Distortion Optimization (RDO), since a higher quality can be attained by adding additional bits. On the basis of these progressive analyzer, Baig \textit{et al.} introduce an inpainting scheme that exploits spatial coherence exhibited by neighboring blocks to reduce redundancy in image \cite{baig2017learning}.

Ball\'e \textit{et al.} relaxed discontinuous quantization step with additive uniform noise to alleviate the non-differentiability, and developed an effective non-linear transform coding framework in the context of compression \cite{balle2016end}. Similarly, Theis \textit{et al.} replaced quantization with a smooth approximation and optimized a convolutional auto-encoder with an incremental training strategy \cite{theis2017lossy}.
Instead of optimizing for pixel fidelity (e.g., Mean Square Error) as most codecs do, Chen \textit{et al.} optimized coding parameters for minimizing semantic difference between the original image and the compressed one \cite{chen2018learning}.

Compared to image, video contains highly temporal correlation between frames. Several works focus on frame interpolation \cite{santurkar2017generative} or frame extrapolation \cite{mathieu2015deep,lotter2016deep,jin2018augmented} to leverage this correlation and increase frame rate. On the other hand, some efforts have been made to estimate optical flow between frames with \cite{dosovitskiy2015flownet} or without \cite{ren2017unsupervised} supervision as footstone for early-stage video analysis. Moreover, Spatial Transformer Networks (STN) \cite{jaderberg2015spatial,sonderby2015recurrent} also capture temporal correlation and provide the ability to spatially transform feature maps by applying parametric transformation to blocks of feature map, allowing it to zoom, rotate and skew the input. However, the quality of synthesis images in these methods is not high enough to be directly applied in video coding.

Encouraged by the aforementioned achievements, we take advantage of the successful exploration in learning-based image compression and further explore a learning-based framework for video compression.

\begin{figure*}[t]
	\centerline{\includegraphics[width=\textwidth]{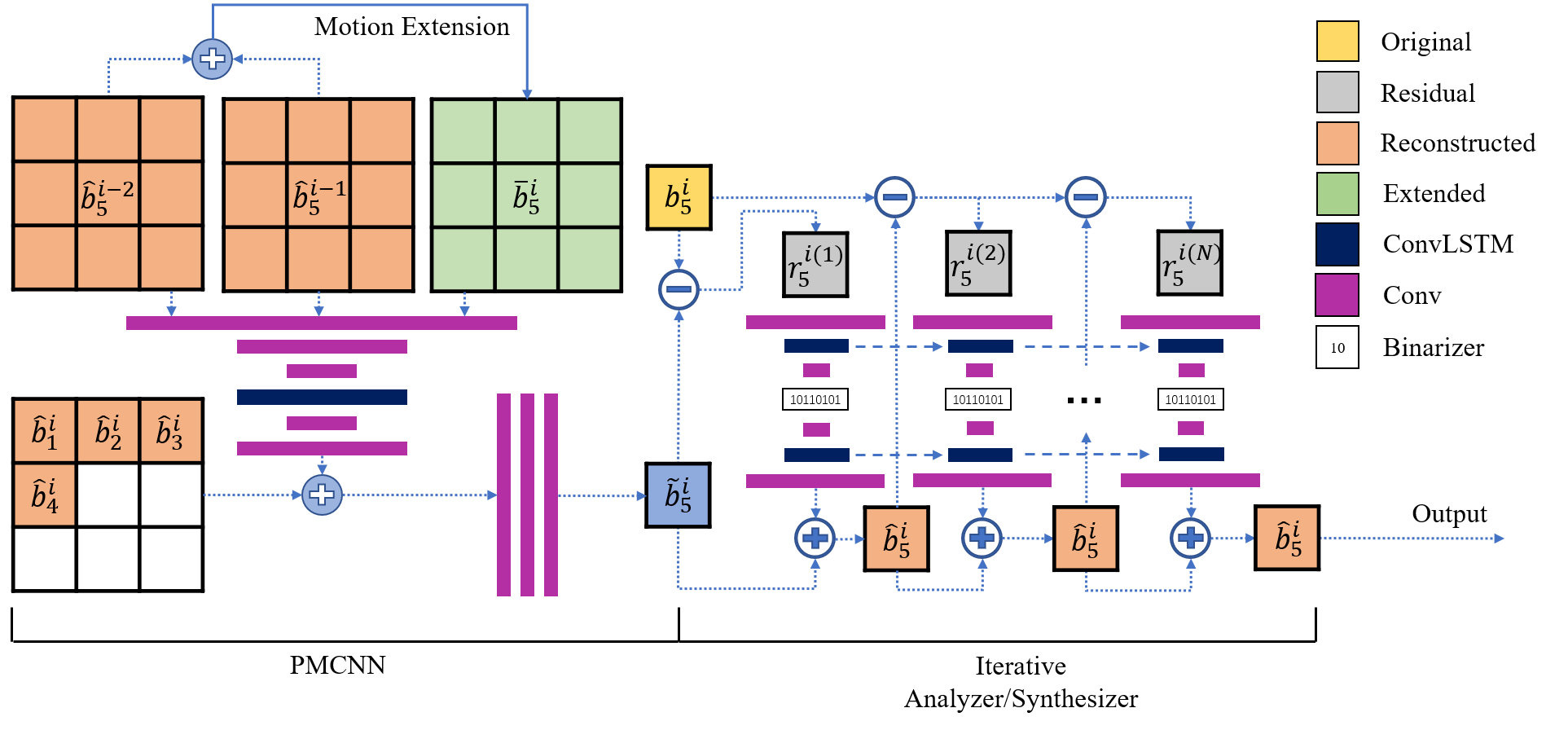}}
	\caption{Detailed architecture of our video compression scheme.}
	\centering
	\label{fig:fig3}
\end{figure*}

\section{Spatiotemporal Modeling with PixelMotionCNN}
\label{spatiotemporal}

In a typical scene, there are great spatiotemporal correlations between pixels in a video sequence. One effective approach to de-correlate highly correlated neighboring signal samples is to model the spatiotemporal distribution of pixels in the video. Since videos are generally encoded and decoded in a sequence, the modeling problem can be solved by estimating a product of conditional distributions (conditioned on reconstructed values), instead of modeling a joint distribution of the pixels. Oord \textit{et al.} \cite{oord2016pixel} introduced PixelCNN for generative image modeling, which can be regarded as a kind of intra-frame prediction for de-correlating neighboring pixels inside one frame. To further de-correlate neighboring frames of the same sequence, we here propose PixelMotionCNN (PMCNN) that sequentially predicts the blocks in a video sequence.

\subsection{Model}
\label{model}

We consider the circumstance where videos are encoded and decoded frame-by-frame in chronological order, and block-by-block in a raster scan order. We define a video sequence $\{f^1, f^2, ..., f^I\}$ as a collection of $m$ frames that are ordered along the time axis. Each frame comprises $n$ blocks sequentialized in a raster scan order, formulated as $f^i=\{b_1^i, b_2^i, ..., b_J^i \}$. Inspired by PixelCNN, we can factorize the distribution of each frame:
\begin{equation}\label{1}
p(f^i) = \prod_{j=1}^J p(b_j^i \mid f^1, ..., f^{i-1}, b_1^i, ..., b_{j-1}^i)
\end{equation}
where $p(b_j^i \mid f^1, ..., f^{i-1}, b_1^i, ..., b_{j-1}^i)$ is the probability of the $j^{th}$ block $b_j^i$ in the $i^{th}$ frame, given the previous frames $f^1, ..., f^{i-1}$ as well as the blocks $b_1^i, ..., b_{j-1}^i$ above and to the left of the current block.

Note that, after the transmission of bitstream, we only have access to the reconstructed data instead of original data at decoding stage. Therefore, as Figure \ref{fig:fig2} illustrates, we sequentially analyze and synthesize each block, saving all the reconstructed frames $\{\hat{f}^1, \hat{f}^2, ..., \hat{f}^I \}$ and blocks $\{\hat{b}_1^i, \hat{b}_2^i, ..., \hat{b}_J^i \}$ in Spatiotemporal Rec-Memory. Instead of conditioning on the previously generated content as PixelCNN does, PMCNN learns to predict the conditional probability distribution $p(b_j^i \mid \hat{f}^1, ..., \hat{f}^{i-1}, \hat{b}_1^i, ..., \hat{b}_{j-1}^i)$ conditioned on previously reconstructed content.

Spatiotemporal modeling is essential to our learning-based video compression pipeline, as it offers excellent de-correlation capability for sources. The experiments in Section \ref{experimental} demonstrate its effectiveness compared with pure spatial or temporal modeling schemes.

\subsection{Architectural Components}

Considering motion information in the temporal direction, we assume that the pixels in neighboring frames are correlated along motion trajectory and there usually exists a linear or non-linear displacement between them. Therefore, as illustrated in Figure \ref{fig:fig3}, we first employ motion extension for approximating linear part of motion displacement, yielding an extended frame $\bar{f}^i$. The non-linear part of motion displacement and the reconstructed blocks (above and to the left of current block) are then jointly modeled with convolutional neural network. This scheme leverages spatiotemporal coherence simultaneously to provide a prediction progressively, and reduces the amount of information required to be transmitted without any other side information (e.g., motion vector).

\textbf{Motion Extension} The objective of motion extension is to extend motion trajectory obtained from previous two reconstructed frames $\hat{f}^{i-2}, \hat{f}^{i-1}$. Let $v_x, v_y, x, y \in \mathbb{Z}$, we first determine a motion vector $(v_x, v_y )$ between $\hat{f}^{i-2}$ and $\hat{f}^{i-1}$ by block matching with $4 \times 4$ block size.
We fill in the whole frame $\hat{f}^{i}$ by copying blocks from $\hat{f}^{i-1}$ according to motion trajectory estimated from corresponding block in $\hat{f}^{i-1}$.
For instance, the values of block $\bar{b}^i$ centered in $(x, y)$ in extended frame $\bar{f}^i$ are copied from $\hat{b}^{i-1}$ centered in $(x-v_x, y-v_y)$. We repeat this operation to obtain complete values of $\bar{f}^i$ as extended frame. It is important to note that there are two intrinsical differences between motion extension and motion estimation \cite{chen2006fast} used in traditional video coding schemes:

\begin{itemize}
	\item We employ motion extension as preprocessing to generate an extended input of PMCNN which utilizes former reconstructed reference frames to generate current coding block. This is essentially different from motion estimation used in traditional codecs which utilizes current coding block to search matching block in reconstructed reference frames.
	\item In general, traditional codecs transmit motion vectors as side information since they indicate where the estimation of current coding block is directly from. By contrast, our proposed scheme doesn't need to transmit motion vectors.
\end{itemize}

Our scheme can definitely obtain a complete frame without gaps since generally motion trajectories is non-linear between frames and can easily be optimized by neural networks.

\begin{figure}[t]
	\begin{subfigure}{0.35\textwidth}
		\includegraphics[width=\linewidth]{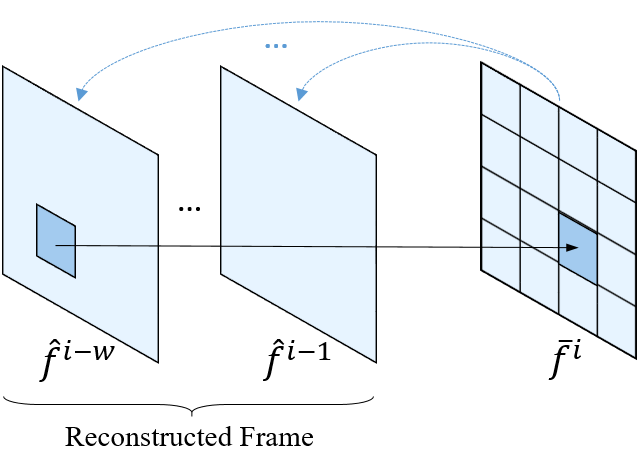}
		\captionsetup{font={footnotesize}}
		\caption{Motion Estimation} \label{fig:fig4a}
	\end{subfigure}
	\begin{subfigure}{0.35\textwidth}
		\includegraphics[width=\linewidth]{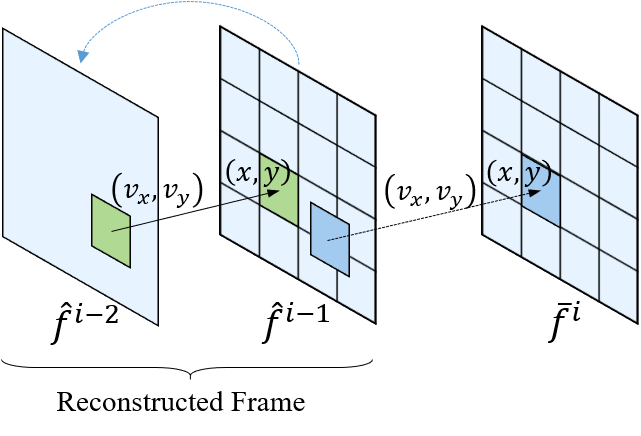}
		\captionsetup{font={footnotesize}}
		\caption{Motion Extension} \label{fig:fig4b}
	\end{subfigure}
	\centering
	\caption[font=footnotesize]{Comparison between motion estimation and motion extension.
		The query frame (the start point of blue dashed arrow) is divided into blocks, and each of the blocks is compared with the blocks in the former frames (the end point of blue dashed arrow) to form a motion vector (the black arrow).
		The black dashed arrow in (b) has the same value as the black arrow, which direct where should the values in $\bar{f}^i$ be copied from.
		Note that, the calculation of motion vectors in (b) only depends on reconstructed content, resulting the omission of transmission, which differs from (a) that transmits motion vectors as side information.
	} \label{fig:fig4}
\end{figure}

\textbf{Hybrid Prediction} In particular, we employ a convolutional neural network that accepts extended frame as its input and outputs an estimation of current block. Previous works \cite{xingjian2015convolutional} have shown that ConvLSTM has the potential to model temporal correlation while reserving spatial invariance. Moreover, residual learning (Res-Block) \cite{he2016deep} is a powerful technique proposed to train very deep convolutional neural network. We here exploit the strength of ConvLSTM and Res-Block to sequentially connect features of $\hat{f}^{i-2}$, $\hat{f}^{i-1}$ and $\bar{f}^i$. An estimation of current frame, as well as the blocks above and to the left of current block, is then fed into several Convolution-BatchNorm-ReLU modules \cite{ioffe2015batch}. As described in Section \ref{model}, the spatiotemporal coherence is modeled concurrently in this scheme, producing a prediction $\tilde{b}_j^i$ of current block.

In this section, we define the form of PMCNN and then describe the detailed architecture of PMCNN \footnote{We give all parameters in the Appendix \ref{appendix:a}.}. In the next section, we give a comprehensive explanation of our framework that employ PMCNN as predictive coding.

\section{Learning for Video Compression}
\label{learning}

Our scheme for video compression can be divided into three components: predictive coding, iterative analysis/synthesis and binarization. Note that entropy coding is not employed in this paper, even though it is commonly applied in standard video compression codecs, and may give more performance gain. In this section, we give details about each component in our scheme and introduce various modes used in our framework.

\textbf{Predictive Coding.} We utilize PMCNN for predictive coding, to create a prediction of a block $\tilde{b}_j^i$ of the current frame based on previously encoded frames as well as the blocks above and to the left of it. This prediction is subtracted from original value to form a residual $r_j^i$. A successful prediction may decrease the energy in the residual compared with original frame, and the data can be represented with fewer bits \cite{o1966predictive}. Our learning objective for the PMCNN can be defined as follows:
\begin{equation}\label{2}
L_{vcnn} = \frac{1}{B \times J} \sum_{i=1}^{B} \sum_{j=1}^{J} (b_j^i - \tilde{b}_j^i)^2,
\end{equation}
where $B$ is batch size, $J$ is the total number of blocks in each frame, and $\tilde{b}_j^i$ denote the output of PMCNN, the superscript and subscript refer to $j^{th}$ block in the $i^{th}$ frame respectively.

\textbf{Iterative Analysis/Synthesis.} Several works put efforts on directly encoding pixel values \cite{toderici2016full,balle2016end}. By contrast we encode the difference between the predicted and the original pixel values. We adopt the model of Toderici \textit{et al.} \cite{toderici2016full}, which is composed of several LSTM-based auto-encoders with connections between adjacent stages. The residuals between reconstruction and target are analyzed and synthesized iteratively to provide a variable-rate compression. Each stage $n$ produces a compact representation required to be transmitted of input residual $r_j^{i(n)}$. We can represent the loss function of iterative analysis/synthesis as follows:
\begin{equation}\label{3}
L_{res} = \frac{1}{B \times J} \sum_{i=1}^{B} \sum_{j=1}^{J} (r_j^{i(1)} - \sum_{m=1}^{S} \hat{r}_j^{i(m)} )^2,
\end{equation}
where
\begin{equation}\label{4}
r_j^{i(1)}=b_j^i - \tilde{b}_j^i,
\end{equation}
\begin{equation}\label{5}
r_j^{i(n)} = r_j^{i(1)} - \sum_{m=1}^{n-1} \hat{r}_j^{i(m)},
\end{equation}
and $\hat{r}_j^{i(n)}$ indicates the output of $n^{th}$ stage, $S$ is the total number of stages ($8$ in this paper).

\begin{figure*}[t]
	\begin{subfigure}{0.18\textwidth}
		\includegraphics[width=\linewidth]{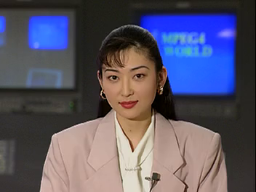}
		\captionsetup{font={footnotesize}}
		\caption{Akiyo} \label{fig:fig5a}
	\end{subfigure}
	\hspace{0.1cm}
	\begin{subfigure}{0.18\textwidth}
		\includegraphics[width=\linewidth]{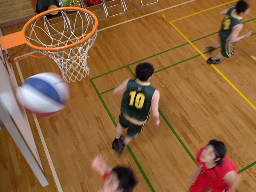}
		\captionsetup{font={footnotesize}}
		\caption{BasketballDrill} \label{fig:fig5b}
	\end{subfigure}
	\hspace{0.1cm}
	\begin{subfigure}{0.18\textwidth}
		\includegraphics[width=\linewidth]{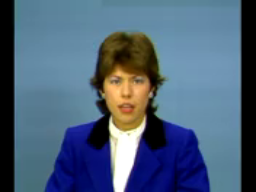}
		\captionsetup{font={footnotesize}}
		\caption{Claire} \label{fig:fig5c}
	\end{subfigure}
	\hspace{0.1cm}
	\begin{subfigure}{0.18\textwidth}
		\includegraphics[width=\linewidth]{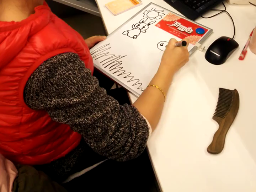}
		\captionsetup{font={footnotesize}}
		\caption{Drawing} \label{fig:fig5d}
	\end{subfigure}
	\hspace{5cm}
	\begin{subfigure}{0.18\textwidth}
		\includegraphics[width=\linewidth]{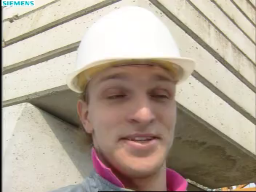}
		\captionsetup{font={footnotesize}}
		\caption{Foreman} \label{fig:fig5e}
	\end{subfigure}
	\hspace{0.1cm}
	\begin{subfigure}{0.18\textwidth}
		\includegraphics[width=\linewidth]{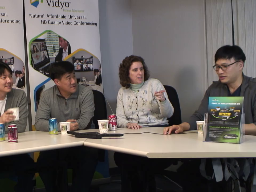}
		\captionsetup{font={footnotesize}}
		\caption{FourPeople} \label{fig:fig5f}
	\end{subfigure}
	\hspace{0.1cm}
	\begin{subfigure}{0.18\textwidth}
		\includegraphics[width=\linewidth]{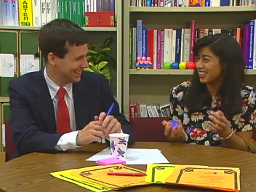}
		\captionsetup{font={footnotesize}}
		\caption{Pairs} \label{fig:fig5g}
	\end{subfigure}
	\hspace{0.1cm}
	\begin{subfigure}{0.18\textwidth}
		\includegraphics[width=\linewidth]{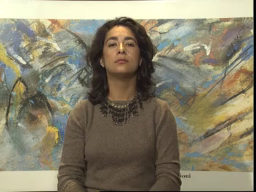}
		\captionsetup{font={footnotesize}}
		\caption{Silent} \label{fig:fig5h}
	\end{subfigure}
	\centering
	\caption{Thumbnail of test sequences.} \label{fig:fig5}
\end{figure*}

\textbf{Binarization.} Binarization is actually where significant amount of data reduction can be attained, since such a many-to-one mapping reduce the number of possible signal values at the cost of introducing some numerical errors. Unfortunately, binarization is an inherently non-differentiable operation that cannot to be optimized with gradient-based techniques. However, some researchers have tackled this problem by mathematical approximation \cite{theis2017lossy,balle2016end,toderici2016full}. Following Raiko \textit{et al.} \cite{raiko2014techniques} and Toderici \textit{et al.} \cite{toderici2015variable}, we add a probabilistic quantization noise $\epsilon$ for the forward pass and keep the gradients unchanged for the backward pass:

\begin{eqnarray}
\epsilon &=&
\left\{
\begin{array}{lll}
1-c_{in}, & \text{with probability} \frac{1+c_{in}}{2} \\
-c_{in}-1, & \text{otherwise,}
\end{array}
\right.
\end{eqnarray}
where $c_{in} \in [-1,1]$ represents the input of binarizer. The output of binarizer can thus be formulated as $c_{out} = c_{in} + \epsilon$. Note that, the binarizer takes no parameter in our scheme.

At test phase, in order to generate extended frame, we need to encode the first two frames directly (without PMCNN). Similarly, we encode the first row and the first column of blocks in each frame only conditioned on previous frames $\{\hat{f}^1, ..., \hat{f}^{i-1} \}$ since they have no spatial neighborhood to be used for predication.

\subsection{Objective Function}

Although there exist potential to use various metrics in this framework for gradient-based optimization, we here employ MSE as the metric for simplicity. During the training phase, PMCNN, iterative analyzer / synthesizer and binarizer are jointly optimized to learn a compact representation of input video sequence. The overall objective can be formulated as:
\begin{equation}\label{6}
L_{total} = L_{vcnn} + L_{res},
\end{equation}
where $L_{vcnn}$ and $L_{res}$ represent the learning objective of PMCNN and iterative analysis/synthesis respectively.

\subsection{Spatially Progressive Coding}

We design a spatially progressive coding scheme in the test phase, by performing various number of iterations determined for each block by a quality metric (e.g. PSNR). This spatially progressive coding scheme enables the functionality of adaptively allocating different bits to different blocks, which can be applied to further improving coding performance similar to rate control in traditional video coding framework. In this paper, we perform this spatially progressive coding scheme in the simplest way, that is to continue to progressively encode residual when the MSE between reconstructed block and the original block is lower than a threshold.

\subsection{Temporally Progressive Coding}

In addition to spatially progressive coding, we also employ a temporally progressive coding scheme that exploits invariance between adjacent frames to progressively determine coding methods of blocks in each frame. We can achieve this by optionally encoding each block according to a specific metric. Also, we implement this scheme in a simplest way in this paper, similar to skip mode defined in traditional video coding framework \cite{sullivan2005video}. In particular, for each block, we first determine whether the block should be encoded or not by calculating MSE between $b_j^i$ and $b_j^{i-1}$. The encoding process is ignored if the MSE is lower than a threshold and a flag (a bit) is transmitted to decoder for indicating the selected mode. The flag is encoded by arithmetic coding as the only overhead ($<1\%$ of bitstream) in our framework. Note that, the encoding of flag has no effect on the training of entire model since it is an out-loop operation.

\begin{figure*}[!t]
	\centerline{\includegraphics[width=14cm]{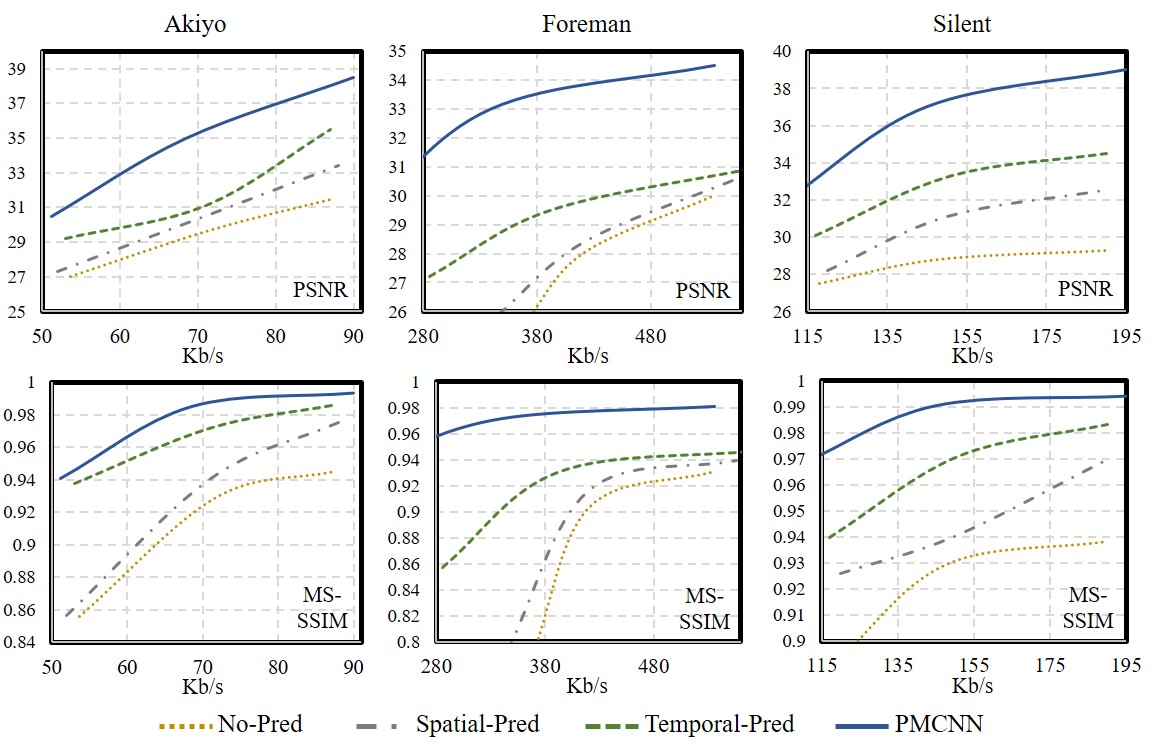}}
	\caption{Quantitative analysis of PMCNN. Each column represents the PSNR/MS-SSIM performance on test sequence. We can observe that PMCNN leverages spatiotemporal dependencies, and surpass the performance of other prediction schemes.}
	\centering
	\label{fig:fig6}
\end{figure*}

\section{Experimental Results}
\label{experimental}

In this section, we first evaluate the performance of PMCNN, then we compare our video compression scheme with traditional video codecs.

\textbf{Dataset.} In general, adjacent frames of the same sequence is highly correlated, resulting in limited diversity. Therefore, we first collect an image dataset for the pre-training of iterative analyzer / synthesizer module as it is used to compress residuals between the reconstructed frame and target frame. The image dataset contains $530,000$ color images collected from Flickr. Each image is down-sampled to 256x256 to enhance the texture complexity. We further perform data augmentation including random rotation and color perturbation during training. Secondly, to model the spatiotemporal distribution of pixels in video sequences, we further jointly train PMCNN module (initialized) and iterative analyzer / synthesizer module (pre-trained) on video dataset, which contains 10,000 sequences sampled from UCF-101. For test set, we collect 8 representative sequences from MPEG/VCEG common test sequences \cite{common2013test} as demonstrated in Figure \ref{fig:fig6}, including various content categories (e.g. global motion, local motion, different motion amplitude, different texture complexity, etc.).
In line with the image dataset, all sequences are resized to 256x192 according to 4:3 aspect ratio. Note that, our video compression scheme is totally block based (fixed 32x32 in our paper), including PMCNN that sequentially predicts blocks and iterative analyzer / synthesizer that progressively compress the residuals between reconstruction and target. Therefore, it can be easily extended to high-resolution scenario.

\textbf{Implementation Details.} We adopt a 32x32 block size for PMCNN and iterative analyzer / synthesizer in our paper according to the verification and comparison of the preliminary experiment. Although variable block size coding typically demonstrates higher performance than fixed block size in traditional codec \cite{vaisey1987variable}, we just verify the effectiveness of our method with fixed block size for simplicity.
We first pretrain PMCNN on aforementioned video dataset using Adam optimizer \cite{kingma2014adam} with $10^{-3}$ learning rate and trained with 20 epochs. The learning rate is decreased by $10^{-1}$ every 5 epochs. All parameters are initialized by Kaiming initialization \cite{he2015delving}.
Our iterative analyzer / synthesizer model is pretrained on 32x32 image blocks randomly cropped from collected image dataset using Adam optimizer with $10^{-3}$ learning rate. We train iterative analyzer / synthesizer with 40 epochs, dropping learning rate by $10^{-1}$ every 10 epochs.
The pretrained PMCNN and pretrained iterative analyzer / synthesizer are then merged and tuned on video dataset with $10^{-4}$ learning rate and trained with 10 epochs. Since block-based video compression framework will inevitably introduce block artifacts, we also implement the simplest deblocking technique \cite{list2003adaptive} to handle block artifacts.

\begin{table}[b]
	\centering
	\caption{Equivalent bit-rate savings (based on PSNR) of different learning-based prediction modes with respect to No-Pred mode.}
	\label{bdrate1}
	\begin{tabular}{lcl|c|l|c|l}
		\toprule[2pt]
		\multicolumn{1}{c}{\multirow{2}{*}{Sequences}} & \multicolumn{2}{c|}{\multirow{2}{*}{Spatial-Pred}} & \multicolumn{2}{c|}{\multirow{2}{*}{Temporal-Pred}} & \multicolumn{2}{c}{\multirow{2}{*}{PMCNN}} \\
		\multicolumn{1}{c}{}                           & \multicolumn{2}{c|}{}                            & \multicolumn{2}{c|}{}                            & \multicolumn{2}{c}{}                          \\ \hline
		\rule{0pt}{3ex}
		Akiyo                                          & \multicolumn{2}{c|}{-9.57\%}                      & \multicolumn{2}{c|}{-42.56\%}                    & \multicolumn{2}{c}{-66.31\%}                  \\
		\rule{0pt}{3ex}
		Foreman                                        & \multicolumn{2}{c|}{-3.82\%}                      & \multicolumn{2}{c|}{-18.47\%}                    & \multicolumn{2}{c}{-56.60\%}                  \\
		\rule{0pt}{3ex}
		Silent                                         & \multicolumn{2}{c|}{-12.12\%}                     & \multicolumn{2}{c|}{-53.23\%}                    & \multicolumn{2}{c}{-72.84\%}                  \\
		\rule{0pt}{3ex}
		Average                                        & \multicolumn{2}{c|}{-6.78\%}                      & \multicolumn{2}{c|}{-30.06\%}                    & \multicolumn{2}{c}{-57.10\%}                  \\
		\bottomrule[1.5pt]
	\end{tabular}
\end{table}

\begin{table}[t]
	\centering
	\caption{Equivalent bit-rate savings (based on PSNR) of Our Scheme with respect to modern codecs.}
	\label{bdrate2}
	\begin{tabular}{lc|l|c|l}
		\toprule[2pt]
		\multicolumn{1}{c}{\multirow{2}{*}{Sequences}} & \multicolumn{2}{c|}{\multirow{2}{*}{MPEG-2}} & \multicolumn{2}{c}{\multirow{2}{*}{H.264}} \\
		\multicolumn{1}{c}{}                           & \multicolumn{2}{c|}{}           & \multicolumn{2}{c}{}           \\ \hline
		\rule{0pt}{3ex}
		Akiyo                                          & \multicolumn{2}{c|}{-52.57\%}                & \multicolumn{2}{c}{-6.12\%}                \\
		\rule{0pt}{3ex}
		BasketballDrill                                & \multicolumn{2}{c|}{-28.31\%}                & \multicolumn{2}{c}{+31.99\%}               \\
		\rule{0pt}{3ex}
		Claire                                         & \multicolumn{2}{c|}{-58.41\%}                & \multicolumn{2}{c}{-40.25\%}               \\
		\rule{0pt}{3ex}
		Drawing                                        & \multicolumn{2}{c|}{-30.41\%}                & \multicolumn{2}{c}{+56.71\%}               \\
		\rule{0pt}{3ex}
		Foreman                                        & \multicolumn{2}{c|}{-39.73\%}                & \multicolumn{2}{c}{+56.67\%}               \\
		\rule{0pt}{3ex}
		FourPeople                                     & \multicolumn{2}{c|}{-80.01\%}                & \multicolumn{2}{c}{-11.71\%}               \\
		\rule{0pt}{3ex}
		Pairs                                          & \multicolumn{2}{c|}{-41.87\%}                & \multicolumn{2}{c}{+8.85\%}                \\
		\rule{0pt}{3ex}
		Silent                                         & \multicolumn{2}{c|}{-56.01\%}                & \multicolumn{2}{c}{-30.74\%}               \\
		\rule{0pt}{3ex}
		Average                                        & \multicolumn{2}{c|}{-48.415\%}               & \multicolumn{2}{c}{+8.175\%}               \\
		\bottomrule[1.5pt]
	\end{tabular}
\end{table}

\begin{figure*}[!t]
	\centerline{\includegraphics[width=18cm]{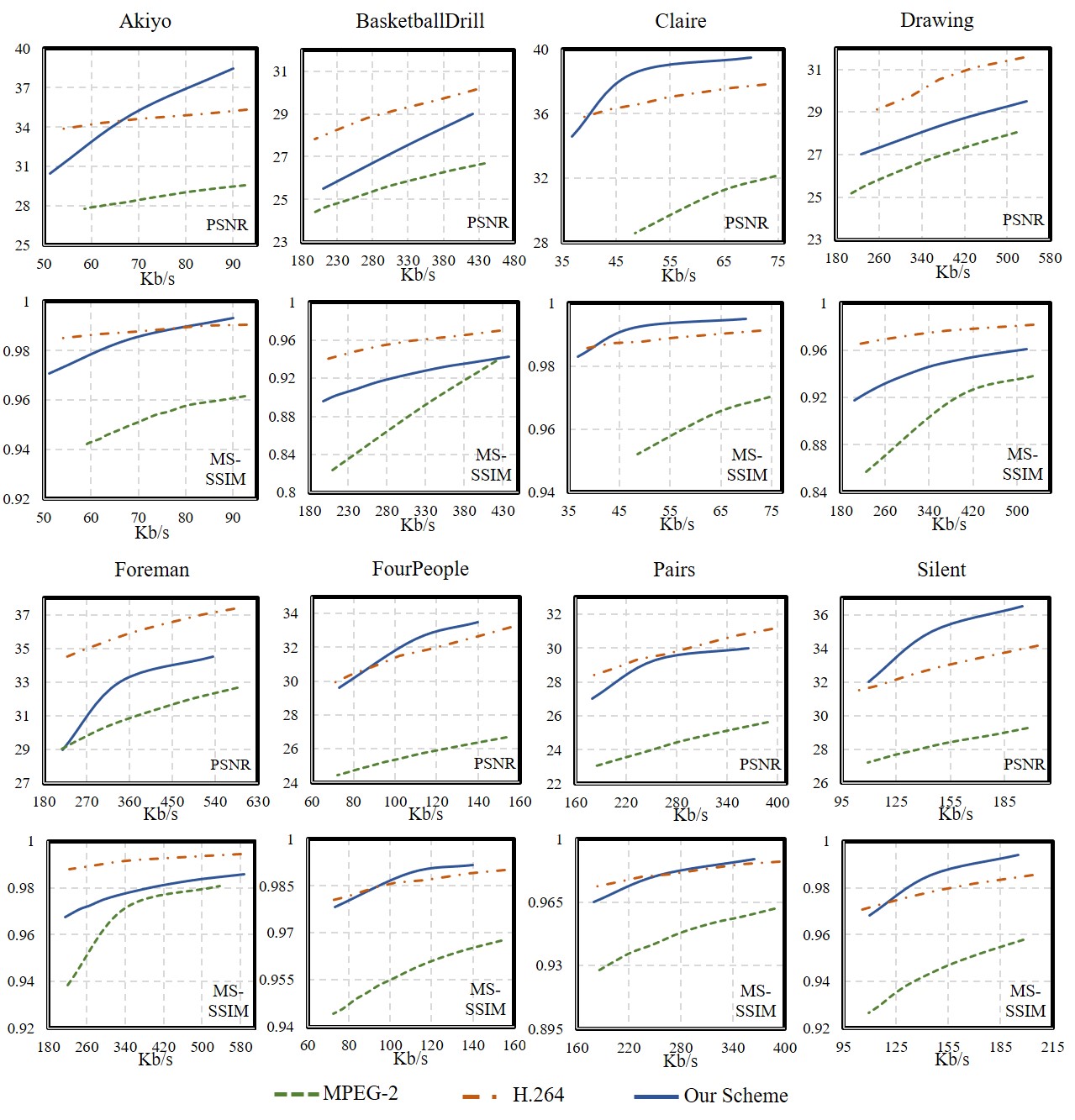}}
	\caption{Quantitative analysis of our learning-based video compression framework. It is worth noting that entropy coding is not employed for our results, even though it is commonly done in standard video compression codecs.}
	\centering
	\label{fig:fig7}
\end{figure*}

\textbf{Baseline.} As the first work of learning-based video compression, we compare our scheme with two representative HVC codecs: MPEG-2 (v1.2) \cite{itu1995generic} and H.264 (JM 19.0) \cite{wiegand2003overview} in our experiments. Both of them take YUV 4:2:0 video format as input and output. We encode the first frame as Intra-frame mode and Predicted-frame mode for the remaining with fixed QP. Rate control is disabled for both codecs.

\textbf{Evaluation Metric.} To assess the performance of our model, we report PSNR between the original videos and the reconstructed ones. Following Toderici \textit{et al.} \cite{toderici2016full}, we also adopt Multi-Scale Structural Similarity (MS-SSIM) \cite{wang2003multiscale} as a perceptual metric. Note that, these metrics are applied both on RGB channels and the reported results in this paper are averaged on each test sequence.

\subsection{Quantitative Analysis of Our Scheme}

\begin{figure*}[h]
	\begin{subfigure}{0.07\textwidth}
		Akiyo
	\end{subfigure}
	\begin{subfigure}{0.165\textwidth}
		\includegraphics[width=\linewidth]{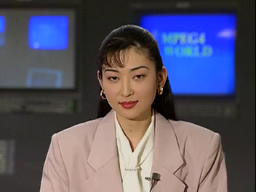}
	\end{subfigure}
	\hspace{0.1cm}
	\begin{subfigure}{0.165\textwidth}
		\includegraphics[width=\linewidth]{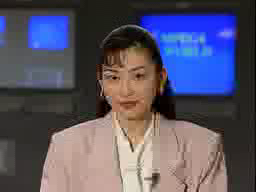}
	\end{subfigure}
	\hspace{0.1cm}
	\begin{subfigure}{0.165\textwidth}
		\includegraphics[width=\linewidth]{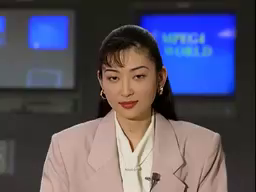}
	\end{subfigure}
	\hspace{0.1cm}
	\begin{subfigure}{0.165\textwidth}
		\includegraphics[width=\linewidth]{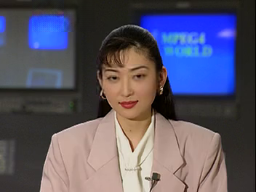}
	\end{subfigure}
	\hspace{5cm}
	\begin{subfigure}{0.07\textwidth}
		Basket-\\
		ballDrill
	\end{subfigure}
	\begin{subfigure}{0.165\textwidth}
		\includegraphics[width=\linewidth]{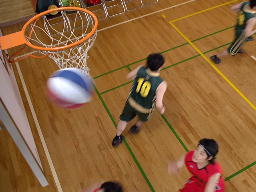}
	\end{subfigure}
	\hspace{0.1cm}
	\begin{subfigure}{0.165\textwidth}
		\includegraphics[width=\linewidth]{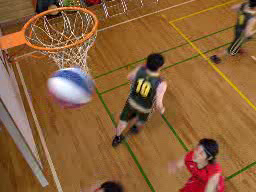}
	\end{subfigure}
	\hspace{0.1cm}
	\begin{subfigure}{0.165\textwidth}
		\includegraphics[width=\linewidth]{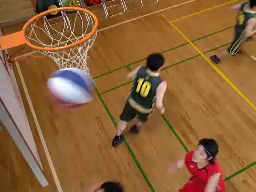}
	\end{subfigure}
	\hspace{0.1cm}
	\begin{subfigure}{0.165\textwidth}
		\includegraphics[width=\linewidth]{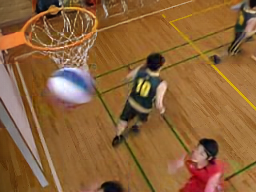}
	\end{subfigure}
	\hspace{5cm}
	\begin{subfigure}{0.07\textwidth}
		Claire
	\end{subfigure}
	\begin{subfigure}{0.165\textwidth}
		\includegraphics[width=\linewidth]{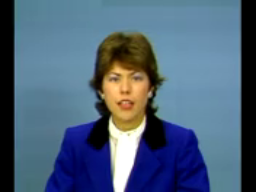}
	\end{subfigure}
	\hspace{0.1cm}
	\begin{subfigure}{0.165\textwidth}
		\includegraphics[width=\linewidth]{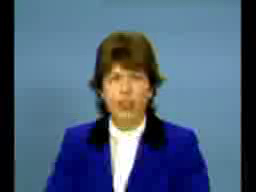}
	\end{subfigure}
	\hspace{0.1cm}
	\begin{subfigure}{0.165\textwidth}
		\includegraphics[width=\linewidth]{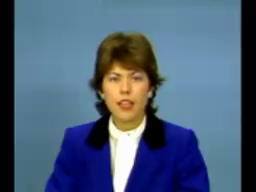}
	\end{subfigure}
	\hspace{0.1cm}
	\begin{subfigure}{0.165\textwidth}
		\includegraphics[width=\linewidth]{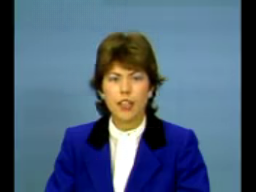}
	\end{subfigure}
	\hspace{5cm}
	\begin{subfigure}{0.07\textwidth}
		Drawing
	\end{subfigure}
	\begin{subfigure}{0.165\textwidth}
		\includegraphics[width=\linewidth]{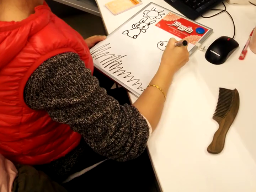}
	\end{subfigure}
	\hspace{0.1cm}
	\begin{subfigure}{0.165\textwidth}
		\includegraphics[width=\linewidth]{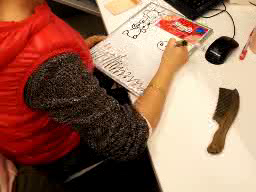}
	\end{subfigure}
	\hspace{0.1cm}
	\begin{subfigure}{0.165\textwidth}
		\includegraphics[width=\linewidth]{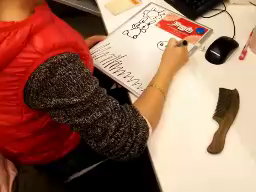}
	\end{subfigure}
	\hspace{0.1cm}
	\begin{subfigure}{0.165\textwidth}
		\includegraphics[width=\linewidth]{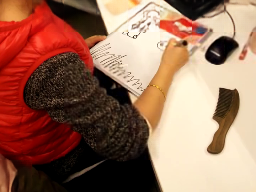}
	\end{subfigure}
	\hspace{5cm}
	\begin{subfigure}{0.07\textwidth}
		Foreman
	\end{subfigure}
	\begin{subfigure}{0.165\textwidth}
		\includegraphics[width=\linewidth]{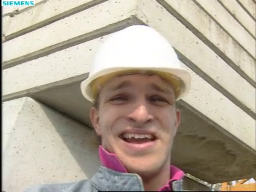}
	\end{subfigure}
	\hspace{0.1cm}
	\begin{subfigure}{0.165\textwidth}
		\includegraphics[width=\linewidth]{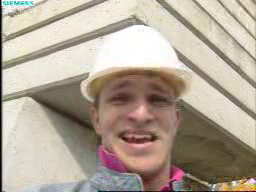}
	\end{subfigure}
	\hspace{0.1cm}
	\begin{subfigure}{0.165\textwidth}
		\includegraphics[width=\linewidth]{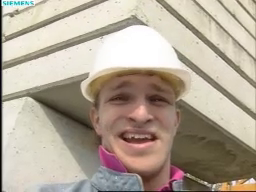}
	\end{subfigure}
	\hspace{0.1cm}
	\begin{subfigure}{0.165\textwidth}
		\includegraphics[width=\linewidth]{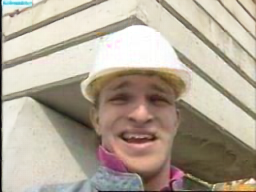}
	\end{subfigure}
	\hspace{5cm}
	\begin{subfigure}{0.07\textwidth}
		Four-\\
		People
	\end{subfigure}
	\begin{subfigure}{0.165\textwidth}
		\includegraphics[width=\linewidth]{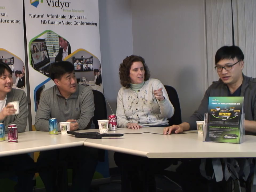}
	\end{subfigure}
	\hspace{0.1cm}
	\begin{subfigure}{0.165\textwidth}
		\includegraphics[width=\linewidth]{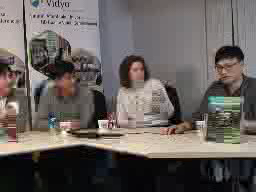}
	\end{subfigure}
	\hspace{0.1cm}
	\begin{subfigure}{0.165\textwidth}
		\includegraphics[width=\linewidth]{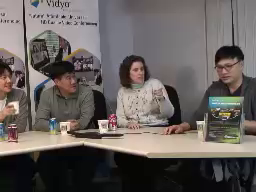}
	\end{subfigure}
	\hspace{0.1cm}
	\begin{subfigure}{0.165\textwidth}
		\includegraphics[width=\linewidth]{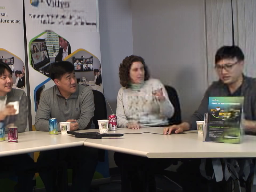}
	\end{subfigure}
	\hspace{5cm}
	\begin{subfigure}{0.07\textwidth}
		Pairs
	\end{subfigure}
	\begin{subfigure}{0.165\textwidth}
		\includegraphics[width=\linewidth]{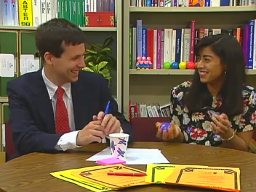}
	\end{subfigure}
	\hspace{0.1cm}
	\begin{subfigure}{0.165\textwidth}
		\includegraphics[width=\linewidth]{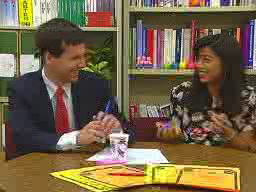}
	\end{subfigure}
	\hspace{0.1cm}
	\begin{subfigure}{0.165\textwidth}
		\includegraphics[width=\linewidth]{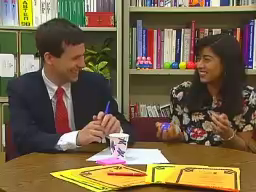}
	\end{subfigure}
	\hspace{0.1cm}
	\begin{subfigure}{0.165\textwidth}
		\includegraphics[width=\linewidth]{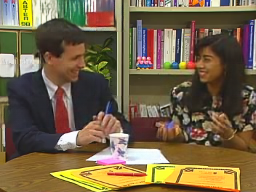}
	\end{subfigure}
	\hspace{5cm}
	\begin{subfigure}{0.07\textwidth}
		Silent
	\end{subfigure}
	\begin{subfigure}{0.165\textwidth}
		\includegraphics[width=\linewidth]{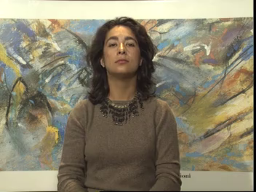}
		\captionsetup{font={footnotesize}}
		\caption{Raw} \label{fig:fig9a}
	\end{subfigure}
	\hspace{0.1cm}
	\begin{subfigure}{0.165\textwidth}
		\includegraphics[width=\linewidth]{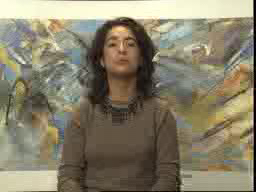}
		\captionsetup{font={footnotesize}}
		\caption{MPEG-2} \label{fig:fig9b}
	\end{subfigure}
	\hspace{0.1cm}
	\begin{subfigure}{0.165\textwidth}
		\includegraphics[width=\linewidth]{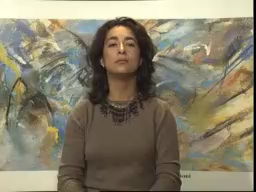}
		\captionsetup{font={footnotesize}}
		\caption{H.264} \label{fig:fig9c}
	\end{subfigure}
	\hspace{0.1cm}
	\begin{subfigure}{0.165\textwidth}
		\includegraphics[width=\linewidth]{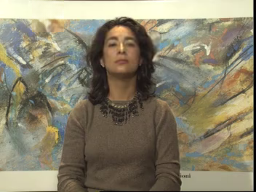}
		\captionsetup{font={footnotesize}}
		\caption{Our Scheme} \label{fig:fig9d}
	\end{subfigure}
	\centering
	\caption{Subjective comparison between various codecs under the same bit-rate.} \label{fig:9}
\end{figure*}

Our proposed PMCNN model leverages spatiotemporal coherence and provide a hybrid prediction $\tilde{b}_j^i$ conditioned on previously reconstructed frames $\hat{f}^1, ..., \hat{f}^{i-1}$ and blocks $\hat{b}_1^i, ..., \hat{b}_{j-1}^i$. To evaluate the impact of each condition, we train our PMCNN conditioned on individual dependency respectively. We refer `Spatial-Pred' as the model trained only conditioned on blocks $\hat{b}_1^i, ..., \hat{b}_{j-1}^i$, `Temporal-Pred' 
as the model trained only conditioned on frames $\hat{f}^1, ..., \hat{f}^{i-1}$, `No-Pred' as the model trained on none of these dependencies. We compare each case with PMCNN on the first 30 frames of three representive sequences including local motion (Akiyo), global motion (Foreman) and different motion amplitude (Silent).

Figure \ref{fig:fig6} demonstrates efficiency of the proposed PMCNN framework, the one simultaneously conditioned on spatial and temporal dependencies (`PMCNN') outperforms the other two patterns that conditioned on individual dependency (`Temporal-Pred' and `Spatial-Pred') or none of these dependencies (`No-Pred'). It is natural because PMCNN modeled on a stronger prior knowledge, while `Temporal-Pred' and `Spatial-Pred' only model the temporal motion trajectory or spatial content relevance respectively.

We further refer BD-rate \cite{bjontegaard2001calcuation} (bit-rate savings) to calculate equivalent bit-rate savings between two compression schemes. As Table \ref{bdrate1} illustrates, compared with individual dependency (`Spatial-Pred' and `Temporal-Pred'), PMCNN effectively exploits spatiotemporal coherence and outperforms `Spatial-Pred' and `Temporal-Pred' significantly.

Our bitstream mainly consists of two parts: the quantized representation generated from iterative analysis / synthesis and flags that indicates the selected mode for temporally progressive coding ($<1\%$ bitstream). Moreover, we employ sequence header, mode header and frame header in bitstream for synchronization. In our experiments, the percentage of skipped blocks is about $25\%\sim89\%$ (influenced by the motion complexity of video content). The Drawing is the sequence with the smallest percentage of skipped blocks, while the Claire achieves the largest. Similar skip detection scheme is also applied in traditional hybrid coding frameworks, with a manner of rate-distortion optimization which is more complex than our proposed scheme. In addition, we replace our LSTM-based analyzer / synthesizer with a series classic convolutional layer (the same number of layers as our scheme). The results indicate that the LSTM-based analyzer / synthesizer can achieve $9.27\%\sim13.21\%$ bit-rate reduction over a convolutional-based analyzer / synthesizer.

We also verify our trained network on three high-resolution sequences without retraining. Compared with H.264 codec, there are around further 8.25\% BD-rate increase by applying the scheme directly to higher resolution sequences. We can expect an improvement by retraining the network, and it's also important to apply variable block size into the scheme, especially for higher resolution content.

\begin{table*}[t]
	\centering
	\caption{Detailed architecture for PMCNN.}
	\label{voxelcnn}
	\begin{tabular}{lllcccclc}
		\toprule[2pt]
		\multicolumn{1}{c}{Layer} & \multicolumn{1}{c}{Type} & \multicolumn{1}{c}{Input}                      & Size / Stride & Dilation  & BN & Activation & \multicolumn{1}{c}{Output} & Output Size \\ \hline
		1                                          & Motion Extension                          & $\hat{f}^{i-2}$, $\hat{f}^{i-1}$                                & -              & -        & -           & -                           & $\bar{f}^i$                                 & 256$\times$192$\times$3      \\
		2                                          & Conv                                      & $\hat{f}^{i-2}$, $\hat{f}^{i-1}$, $\bar{f}^i$                   & 4$\times$4 / 2 & - & Y                   & ReLU                        & conv2                                       & 128$\times$96$\times$96      \\
		3                                          & ResBlock$\times$4                         & conv2                                                           & -     & -         & -                   & -                           & rb3                                         & 128$\times$96$\times$96      \\
		4                                          & Conv                                      & rb3                                                             & 4$\times$4 / 2 & - & Y                   & ReLU                        & conv4                                       & 64$\times$48$\times$192      \\
		5                                          & ResBlock$\times$8                         & conv4                                                           & -      & -        & -                   & -                           & rb5                                         & 64$\times$48$\times$192      \\
		6                                          & Conv                                      & rb5                                                             & 4$\times$4 / 2 & - & Y                   & ReLU                        & conv6                                       & 32$\times$24$\times$192      \\
		7                                          & ResBlock$\times$12                        & conv6                                                           & -      & -        & -                   & -                           & rb7                                         & 32$\times$24$\times$192      \\
		8                                          & Conv                                      & rb7                                                             & 4$\times$4 / 2 & - & Y                   & ReLU                        & conv8                                       & 16$\times$12$\times$96       \\
		9                                          & ConvLSTM                                  & conv8                                                           & 3$\times$3 / 1 & - & N                   & -                           & convlstm9                                   & 16$\times$12$\times$32       \\
		10                                         & DeConv                                    & convlstm9                                                       & 5$\times$5 / 2 & - & Y                   & ReLU                        & deconv10                                    & 32$\times$24$\times$32       \\
		11                                         & Pooling                                   & $\bar{f}^i$                                                     & 5$\times$5 / 8 & - & N                   & -                           & pooling11                                   & 32$\times$24$\times$32       \\
		12                                         & Conv                                      & pooling11                                                       & 4$\times$4 / 1 & - & Y                   & ReLU                        & conv12                                      & 32$\times$24$\times$32       \\
		13                                         & Concat                                    & deconv10, conv12                                                & -              & -       & -            & -                           & concat13                                    & 32$\times$24$\times$64       \\
		14                                         & ResBlock$\times$12                        & concat13                                                        & -              & -           & -        & -                           & rb14                                        & 32$\times$24$\times$64       \\
		15                                         & DeConv                                    & rb14                                                            & 5$\times$5 / 2 & - & Y                   & ReLU                        & deconv15                                    & 64$\times$48$\times$32       \\
		16                                         & Pooling                                   & $\bar{f}^i$                                                     & 5$\times$5 / 4 & - & N                   & -                           & pooling16                                   & 64$\times$48$\times$32       \\
		17                                         & Conv                                      & pooling16                                                       & 4$\times$4 / 1 & - & Y                   & ReLU                        & conv17                                      & 64$\times$48$\times$32       \\
		18                                         & Concat                                    & deconv15, conv17                                                & -              & -       & -            & -                           & concat18                                    & 64$\times$48$\times$64       \\
		19                                         & ResBlock$\times$8                         & concat18                                                        & -              & -         & -          & -                           & rb19                                        & 64$\times$48$\times$64       \\
		20                                         & DeConv                                    & rb19                                                            & 5$\times$5 / 2 & - & Y                   & ReLU                        & deconv20                                    & 128$\times$96$\times$16      \\
		21                                         & Pooling                                   & $\bar{f}^i$                                                     & 5$\times$5 / 2 & - & N                   & -                           & pooling21                                   & 128$\times$96$\times$16      \\
		22                                         & Conv                                      & pooling21                                                       & 4$\times$4 / 1 & - & Y                   & ReLU                        & conv22                                      & 128$\times$96$\times$16      \\
		23                                         & Concat                                    & deconv20, conv22                                                & -              & -        & -           & -                           & concat23                                    & 128$\times$96$\times$32      \\
		24                                         & ResBlock$\times$4                         & concat23                                                        & -              & -          & -         & -                           & rb24                                        & 128$\times$96$\times$32      \\
		25                                         & DeConv                                    & rb24                                                            & 5$\times$5 / 2 & - & Y                   & Tanh                        & deconv25                                    & 256$\times$192$\times$3      \\
		26                                         & Conv                                      & $\bar{f}^i$, deconv25                                           & 4$\times$4 / 1 & - & N                   & Tanh                        & conv26                                      & 256$\times$192$\times$3      \\
		27                                         & ConvBlock$\times$8                        & $\hat{b}_{j-9}^i$, $\hat{b}_{j-8}^i$, $\hat{b}_{j-1}^i$, conv26 & -              & -        & -           & Tanh                           & cb27                                        & 64$\times$64$\times$3        \\
		28                                         & Crop                                      & cb27                                                            & -              & -            & -       & -                           & output                                      & 32$\times$32$\times$3        \\
		\hline
		\specialrule{0em}{1pt}{1pt}
		\multicolumn{9}{c}{ResBlock} \\ \hline
		1                                         & Conv                                      & input                                                            & 3$\times$3 / 1      & -        & Y                   & ReLU                           & conv1                                      & -        \\
		2                                         & Conv                                      & conv1                                                            & 3$\times$3 / 1      & -        & N                   & ReLU                           & conv2                                      & -        \\
		3                                         & Add                                      & input, conv2                                                            & -              & -      & -             & -                           & output                                      & -        \\
		\hline
		\specialrule{0em}{1pt}{1pt}
		\multicolumn{9}{c}{ConvBlock} \\ \hline
		1                                         & DilConv                                      & input                                                            & 3$\times$3 / 1      & 1        & Y                   & ReLU                           & conv1                                      & -        \\
		2                                         & DilConv                                      & input                                                            & 3$\times$3 / 1      & 2        & Y                   & ReLU                           & conv2                                      & -        \\
		3                                         & DilConv                                      & input                                                            & 3$\times$3 / 1      & 4        & Y                   & ReLU                           & conv3                                      & -        \\
		4                                         & DilConv                                      & input                                                            & 3$\times$3 / 1      & 8        & Y                   & ReLU                           & conv4                                      & -        \\
		5                                         & Concat                                      & conv1, conv2, conv3, conv4                                                            & -      & -        & N                   & -                           & output                                      & -        \\
		\bottomrule[1.5pt]
	\end{tabular}
\end{table*}

\subsection{Comparison with Traditional Video Codecs}

As the first work of learning-based video compression, we quantitatively analyze the performance of our framework and compare it with modern video codecs.

We provide quantitative comparison with traditional video codecs in Table \ref{bdrate2} and Figure \ref{fig:fig7}, as well as subjective quality comparison in Figure \ref{fig:9}. The experimental results illustrate that our proposed scheme outperforms MPEG-2 significantly with 48.415\% BD-Rate reduction and correspondingly 2.39dB BD-PSNR \cite{bjontegaard2008improvements} improvement in average, and demonstrates comparable results with H.264 codec with around 8.175\% BD-Rate increase and correspondingly 0.41dB BD-PSNR drop in average. We achieve this despite the facts:

\begin{itemize}
	\item We do not perform entropy coding in our aforementioned experiments since it is not the main contribution of this paper and there should be more research work to design an optimized entropy encoder for learning based video compression framework. In order to demonstrate the potential of this research, we have tried a simple entropy coding method described in \cite{toderici2016full} without any specific optimization, an average performance gain of $12.57\%$ can be obtained compared to our scheme without entropy coding.
	\item We do not perform complex prediction modes selection or adaptive transformation schemes as developed for decades in traditional video coding schemes.
\end{itemize}

Although affected by the above factors and our learning-based video compression framework is in its infancy stage to compete with latest H.265/HEVC video coding technologies\cite{sullivan2012overview}, it still shows great improvement over the first successful video codec MPEG-2 and has enormous potential in the following aspects:

\begin{itemize}
	\item We provide a possible new direction to solve the limitation of heuristics in HVC by learning the parameters of encoder/decoder.
	\item The gradient-based optimization used in our framework can be seamlessly integrated with various metrics (loss function) including perceptual fidelity and semantic fidelity, which is infeasible to HVC. For instance, a neural network-based object tracking algorithm can be employed as a semantic metric for surveillance video compression.
	\item We have less side information required to be transmitted than HVC. The only overhead in our scheme is the flag ($<1\%$ of bitstream) used for temporal progressive coding. By contrast, HVC require considerable side information (e.g., motion vector, block partition, prediction mode information, etc.) to indicate sophisticated coding modes.
\end{itemize}

We calculate the time consuming of our scheme and traditional codecs on the same machine (CPU: i7-4790K, GPU: NVIDIA GTX 1080). The overall computational complexity of our implementation is about 141 times that of H.264 (JM 19.0). It should be noted that our scheme is just a preliminary exploration of learning-based framework for video compression and each part is implemented without any optimization. We believe the overall computational complexity can be reduced in the future by applying algorithm optimization based on specific AI hardware, and some existing algorithms can also be revised accordingly, e.g., some parallel processing like wave front parallelism, some efficient network architecture (e.g., ShuffleNet, MobileNet) or adopting some network model compression techniques (e.g., pruning, distillation).

We also observe that, our approach shows unstable performance on various test sequences (especially in the case of global motion). This is reasonable since the coding modes we adopted in our algorithm are still very simple and unbalanced. However, we successfully demonstrate the potential of this framework and provide a potential new direction for video compression.

\section{Conclusion}
\label{conclusion}

We propose the concept of PMCNN by modeling spatiotemporal coherence to effectively perform predictive coding and explore a learning-based framework for video compression. Although lack of entropy coding, this scheme still achieves a promising result for video compression, demonstrating a new possible direction of video compression. In the future, we will apply our scheme to high-resolution video sequence and expect more optimization since there are lots of aspects can be further improved in this framework, including entropy coding, variable block size coding, enhanced prediction, advanced post-processing techniques and integration of various metrics (e.g., perceptual/semantic metric), etc.

\appendices

\section{Detailed architecture for PMCNN}
\label{appendix:a}
We provide all parameters in PMCNN in the Table \ref{voxelcnn}. The notation is consistent with paper. In addition, `BN' denotes Batch Normalization. `DeConv' denotes deconvolution layer. `Concat' denotes concatenate feature maps along the last dimension. `DilConv' denotes dilated convolution layer.





\ifCLASSOPTIONcaptionsoff
  \newpage
\fi




\bibliographystyle{IEEEtran}
\bibliography{videotrans}
\end{document}